# Third Party Privacy Preserving Protocol for Perturbation Based Classification of Vertically Fragmented Data Bases

B.Hanmanthu[a],a*, B.Raghu Ram[a], P.Niranjan[b]

[a]Assistant Professor, Department of Computer Science and Engineering, KITS, Warangal-506 015, A.P., INDIA.
[b]Professor&Head, Department of Computer Science and Engineering, KITS, Warangal-506 015, A.P., INDIA.

**Abstract**

Privacy is become major issue in distributed data mining. In the literature we can found many proposals of privacy preserving which can be divided into two major categories that is trusted third party and multiparty based privacy protocols. In case of trusted third party models the conventional asymmetric cryptographic based techniques will be used and in case of multi party based protocols data perturbed to make sure no other party to understand original data. In order to enhance security features by combining strengths of both models in this paper, we propose to use data perturbed techniques in third party privacy preserving protocol to conduct the classification on vertically fragmented data bases. Specially, we present a method to build Naive Bayes classification from the disguised and decentralized databases. In order to perform classification we propose third party protocol for secure computations. We conduct experiments to compare the accuracy of our Naive Bayes with the one built from the original undisguised data. Our results show that although the data are disguised and decentralized, our method can still achieve fairly high accuracy.

## 1. Introduction

Distributed data mining [1] playing great role in extracting knowledge from geographically distributed sources. The exposures of distributed data mining include sensor networks, mobile ad hock communications, context aware computing, weather forecasting, intruder detection systems and web mining etc. The corporate and financial sectors as they spread over different geographic locations transforming their business intelligence applications to distributed application form conventional centralized data warehouse applications. Even though centralized data warehouse based models are more accurate than distributed applications, due to their easy scalability, communication efficiency and privacy facilities distributed data mining applications attracting lot of researcher's attention.

Data mining approaches on geo graphically distributed data sources can be applied on two ways know as centralized model and distributed model. In the centralized model the required data distributed over various sources gathered in to a centralized site where mining algorithm will be applied. It provides accurate results but impose huge communication cost and time. Where as in distributed environment, mining will be performed at local sites and results of local sites will be optimized based on feedback from other sites. Even though distributed model will gives less accuracy than centralized model it reduce communication cost, time complexity and makes algorithm easily scalable. In order to develop model with optimum time and communication cost we adopted distributed environment.

In distributed environment data can be handled by horizontal partitioning in which transactions with same attributes distributed among different sites or by vertical partitioning in which each site possess the common set of transaction by distributing attributes among different sites. Even though horizontal portioning applications found in many real time applications, vertical partitioning [2] also have its significance in real time applications for example in bank system customer information related to salary transactions may store at one site and information related to loan transactions may store at different database with same customer id. In some cases even different firms like banks and insurance companies may also want share their customer information to study common patterns. This kind of applications falls in to vertically partitioned applications where with same id different set of attributes stored at different sites and common patterns among such sites need to be extracted. A lot of research work is progressing in transforming conventional centralized data mining applications to handle vertically partitioned databases.

The classification techniques which plays important role in distributed learning with scientific and commercial applications can be implemented throw many techniques include decision trees, regression and Navi's Bayesian based methods etc. Out of all different classification techniques Naive Bayes proved efficient by means of accuracy and interoperability.

In the literature we can found many proposals of privacy preserving which can be divided into two major categories that is trusted third party and multiparty based privacy protocols. In case of trusted third party models the conventional asymmetric cryptographic based techniques will be used and in case of multi party based protocols data perturbed to make sure no other party to understand original data. In order to enhance security features by combining strengths of both models in this paper, we propose to use data perturbed techniques in third party privacy preserving protocol to conduct the classification on vertically fragmented data bases. The remaining part of the paper organized in following manner, in section 2 presents' related works and in section 3 naïve bayes classification model for vertically fragmented data is given finally the performance and evaluation given in section 4.

## 2. Related Work

Previous work in privacy-preserving data mining has addressed two issues. In one, the aim is to preserve customer privacy by perturbing the data values [3]. In this scheme random noise data is introduced to distort sensitive values, and the distribution of the random data is used to generate a new data distribution which is close to the original data distribution without revealing the original data values. The estimated original data distribution is used to reconstruct the data, and data mining techniques, such as classifiers and association rules are applied to the reconstructed data set. Later refinement of this approach has tightened estimation of original values based on the distorted data [4]. Perturbation methods and their privacy protection have been criticized because some methods may derive private information from the reconstruction step [5]. One important category is multiplicative perturbation method. In the view of geometric property of the data, multiplying the original data values with a random noise matrix is to rotate the original data matrix, so it is also called rotated based perturbation.

A number of other techniques [6] [7] have also been proposed for privacy preservation which works on different classifiers such as in [6] combine the two of data transform and data hiding to propose a new randomization method, Randomized Response with Partial Hiding (RRPH), for distorting the original data. Then, an effective Naive Bayes classifier is presented to predict the class labels for unknown samples according to the distorted data by RRPH. In [7],Proposes optimal randomization schemes for privacy preserving density estimation. The work in [8] [5] describes the methods of improving the effectiveness of classification such as in [8], it proposes two algorithms BiBoost and MultBoost which allow two or more participants to construct a boosting classifier without explicitly sharing their data sets and analyze both the computational and the security aspects of the algorithms. In case of distributed environment, the most widely used technique in privacy preservation mining is secure sum computation [9].

Privacy-preserving algorithms on vertically partitioned data have been proposed with different techniques including association rules mining [10] and classification [11], [12]. These privacy solutions can be broadly categorized into two approaches. One approach adopts cryptographic techniques to provide secure solutions in distributed settings [13]. Another approach randomizes the original data in such a way that hides the underlying patterns [14] which can affect quality of results.

To our knowledge, very few works focus on mapping or modifying perturbation technique with third party based protocol. In order to provide enhanced security feature we propose trusted third party based naive bays classifier of perturbed data on vertically fragmented databases.

## 3. Trusted Third Party Privacy Preserving Protocol for Distributed Naïve Bayes classification of Perturbed Data

In the distributed third-party setting as shown in "Fig.1" a database set D hold confidential databases D1,D2…..Dn, respectively, each of which can be regarded as a relational table. Each database has same number rows. All this sub databases of D shares the sub set of variable of same transactions. There is a common ID that links the rows in distributed in among sites and all subset of transactions also holds the class label to which transaction belongs. A trusted third party data miner initiates the associative classification rules extraction process by broadcasting minimum support count. All sites generate Naive Bayes statistical values for each class label. The locally generated values and private key encrypted attribute vector send to trusted third party. All local sites will repeat the same procedure. The values send to trusted third party which holds public key of all sites to generate naive's Bays classification rules.

Our model aimed to perform Naive Bayes based classification on distributed data bases where data stored in various shred nothing machine connected over distributive environment. Our model assumes that data is vertically distributed over various sources. In distributive environment we may face two criteria's out of one is the training data distributed among various sources and classification as to perform on one source. In other approach along with training data distribution the classification also need to perform on various sources. We proposed our model to handle both of these factors. The proposed distributed decision tree based classifier model will follow three steps in order to perform associative classification on distributed data bases.

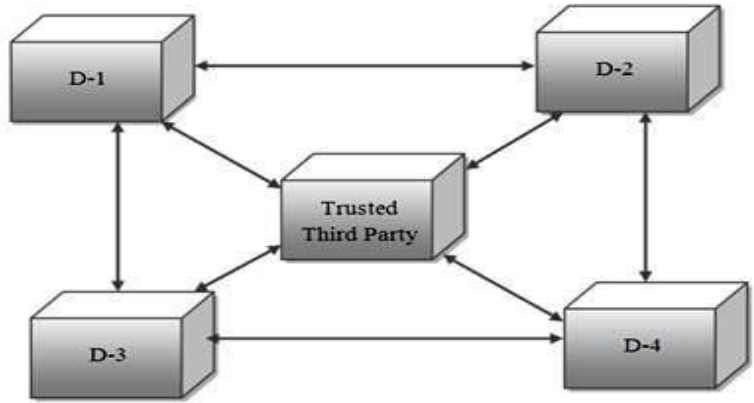

Fig. 1.Trusted Third Party Based System for Classification;

## 3.1. Trusted Third Party Protocol for Perturbed Data

Step1: The process will be started by the trusted third party by broadcasting mining initiation message.
Step2: All sites which ever interest will respond by sending ready message and their public keys.
Step3: Then the trusted third party will respond by start message.
Step4: Each site generates its perturbed data and calculates individual local count of instances for Naive Bayes based statistical values.
Step5: The individual sites classified data will encrypt the required data by private key using RSA Algorithm.
Step6: The individual site sends their encrypted data to trusted third party.
Step7: The trusted third party which hold public key will decrypt data and apply perturbed naive bayes classifier constructor as shown in next section.

## 3.2. Naïve Bayes Classifier Construction Over Perturbed Data

As stated in [10], Naive Bayes classifier can be applied directly on the perturbed data. The Naive Bayes classifier labels a new instance by assigning the most probable class value. Besides, it assumes that attribute values are conditionally independent given the class value in order to simplify the estimation of the required probabilities. Using the above assumptions, Naïve Bayes classifier selects the most likely classification *Cnb* as[11]

We need to estimate $\mu_{ij}$ and $\sigma^2_{ij}$ for each attribute $X_i$ and for each class label $C_j$ using the perturbed numeric data to construct a Naive Bayes classifier. In the perturbed data case, instead of the original attribute value $X_i$, we only see the $W_i = X_i + R$ values. Let $wt_{ij}$ be the ith attribute value of the tth training data instance with class label $C_j$ In addition, we assume that there are n instances with class label $C_j$.

We also know that $wt_{ij} = xt_{ij} + rt_{ij}$ where $rt_{ij}$ is the randomly generated noise with mean zero and known variance $\sigma^2_R$. Using the above facts, we can show that the expected value of sample variance since the Sample variance $S^2 = \frac{1}{n-1} \cdot \sum_{t=1}^{n}(w^t_{ij} - w_j)^2$ has an expected value $\sigma^2_{ij} + \sigma^2_R$ we can use $S^2$ and the known $\sigma^2_R$ to estimate the $\sigma^2_{ij}$ (i.e. use $S^2 - \sigma^2_R$ to estimate $\sigma^2_{ij}$).

As a result, as long as we do not change the class labels, we can directly construct Naive Bayes classifier from the perturbed data. Even more, since the parameter estimations done by using the perturbed data and the original data have the same expected values, we should be able to get similar classification accuracy in both cases.

## 4. Performance Analysis

In order to testify the performance of our proposed model, our experiments utilized four P4 2.40GHz PCs with 512Mb main memory and windows XP Operating System. The four PCs are located in 100Mb LAN out one we used as trusted third which initiate the process rest of stores data sets. We use the pima-indians-diabetes_data(PIMA) and Heart disease data sets obtained from UCI Machine Learning Repository.

*4.1. Experiment Setup*

At first step of experiment the data sets are processed by discredited quantitative items and partition data sets vertically among three systems connected in distributed environment. For both PIMA and Heart datasets we created unique id for each row and split in to three parts where each part row carry same id and stored in three different systems. Then data at local sites divided horizontally into training and testing data sets in 50:50 ratios using 10X10 validation method. In second step proposed Naive Bayes algorithm implemented on training data sets at local sites and generated results encrypted using private key and send public key to trusted third party for this method we adopted RSA algorithm. In third step combining all local data of individual sites gathered at trusted third party will used to generate global Naive Bayes classifier. The classifier generated send to all site because our experiment intended to classify data present at all local sites. Finally classifier applied on test data sets without class labels. The classifier efficiency calculated by comparing generated class labels against actual class labels.

*4.2. Performance Evaluation*

In order to evaluate accuracy of our vertically distributed associative classifier we compared obtained accuracy against standard associative classifier models. We adopted support threshold as 20 and confidence threshold as 80 according to standards. This proposed privacy preserving association rules based classification model on vertically portioned databases shown accuracy of 77.2% on test data bases of UCI PIMA dataset and 81.7 on test data base of UCI Heart disease data sets which shows efficiency of our model.

Our proposed model gives lesser communication cost because the data transfers among sites and last site to miner is performed as a single bulk data transfer instead of single data transfer for each frequent item set. So as many numbers of sites are there that many number of data transfers are needed to obtain global frequent item sets.

*4.3. Security Analysis*

Our model preserves privacy of results sent to successor site by a site in order to finding global frequent item sets because local frequent values sent in scalar matrix form and item set names sent in encrypted vector form. This is possible because we are using scalar product method to generate frequent item sets using matrixes of two sites.

The trusted third party is having certain privileges such as initiation of the mining process, decryption of frequent item sets, finding global frequent item sets and association rules but it can't predict any site's private data because the only the global frequent data came to it after processing at local sites.

## 5. Conclusion

In this paper we proposed enhance security feature model by combining strengths of both models in this paper, we propose to use data perturbed techniques in third party privacy preserving protocol to conduct the classification on vertically distributed data bases. Specially, we present a method to build Naive Bayes classification from the disguised and vertically fragmented databases. The experiment conducted on UCI data sets proved that our model shown good accuracy measures. We also showed analytically that our model is time, cost and privacy efficient.